\documentclass[letterpaper,useAMS,usenatbib]{mn2e}

\pdfoutput=1

\usepackage{color}
\usepackage{graphicx}
\usepackage{graphics}
\usepackage{rotating}
\usepackage{amsmath}        
\usepackage{amssymb}        
\usepackage{epsfig}        
\usepackage[T1]{fontenc}
\usepackage{aecompl}
\usepackage{times}

\newcommand{\apjs}{ApJS}

\newcommand{\pasa}{PASA}

\newcommand{\degree}{$^{\circ}$}

\newcommand{\ha}{H$\alpha$}

\title[A BH candidate in 47 Tuc]{Deep radio imaging of 47 Tuc identifies the peculiar X-ray source X9 as a new black hole candidate}

\author[J.C.A.~Miller-Jones et al.]
 {J.C.A.~Miller-Jones,$^{1}$\thanks{email: james.miller-jones@curtin.edu.au}
J.~Strader,$^2$ C.O.~Heinke,$^{3,4}$ T.J.~Maccarone,$^5$ \and M.~van den Berg,$^{6,7}$ C.~Knigge,$^8$ L.~Chomiuk,$^2$ E.~Noyola,$^9$ T.D.~Russell,$^1$ \and A.C.~Seth,$^{10}$ and G.R.~Sivakoff\,$^3$\\
$^1$International Centre for Radio Astronomy Research - Curtin University, GPO Box U1987, Perth, WA 6845, Australia\\
$^2$Department of Physics and Astronomy, Michigan State University, East Lansing, MI 48824, USA\\
$^3$Department of Physics, University of Alberta, Room 238 CEB, Edmonton, AB T6G 2G7, Canada\\
$^4$Alexander von Humboldt Fellow, at Max-Planck-Institut f\"ur Radioastronomie, Auf dem H\"ugel 69, D-53121 Bonn, Germany\\
$^5$Department of Physics, Texas Tech University, Box 41051, Lubbock, TX 79409-1051, USA\\
$^6$Anton Pannekoek Institute for Astronomy, University of Amsterdam, Science Park 904, 1098 XH, Amsterdam, the Netherlands\\
$^7$Harvard-Smithsonian Center for Astrophysics, 60 Garden Street, Cambridge, MA 02138, USA\\
$^8$Physics and Astronomy, University of Southampton, Highfield, Southampton, SO17 1BJ, UK\\
$^9$McDonald Observatory, The University of Texas at Austin, 2515 Speedway, Stop C1402, Austin, TX 78712-1206, USA\\
$^{10}$Department of Physics and Astronomy, University of Utah, 115 South 1400 East, Salt Lake City, Utah 84112, USA\\
}
\begin{document}

\date{Accepted 2015 August 11. Received 2015 August 11; in original form 2015 May 8.}

\pagerange{\pageref{firstpage}--\pageref{lastpage}} \pubyear{2015}

\maketitle

\label{firstpage}

\begin{abstract}
We report the detection of steady radio emission from the known X-ray source X9 in the globular cluster 47 Tuc.  With a double-peaked C\,{\sc iv} emission line in its ultraviolet spectrum providing a clear signature of accretion, this source had been previously classified as a cataclysmic variable.  In deep ATCA imaging from 2010 and 2013, we identified a steady radio source at both 5.5 and 9.0\,GHz, with a radio spectral index (defined as $S_{\nu}\propto\nu^{\alpha}$) of $\alpha=-0.4\pm0.4$.  Our measured flux density of $42\pm4$\,$\mu$Jy\,beam$^{-1}$ at 5.5 GHz implies a radio luminosity ($\nu L_{\nu}$) of $5.8\times10^{27}$\,erg\,s$^{-1}$, significantly higher than any previous radio detection of an accreting white dwarf.  Transitional millisecond pulsars, which have the highest radio-to-X-ray flux ratios among accreting neutron stars (still a factor of a few below accreting black holes at the same $L_X$), show distinctly different patterns of X-ray and radio variability than X9.  When combined with archival X-ray measurements, our radio detection places 47 Tuc X9 very close to the radio/X-ray correlation for accreting black holes, and we explore the possibility that this source is instead a quiescent stellar-mass black hole X-ray binary.  The nature of the donor star is uncertain; although the luminosity of the optical counterpart is consistent with a low-mass main sequence donor star, the mass transfer rate required to produce the high quiescent X-ray luminosity of $10^{33}$\,erg\,s$^{-1}$ suggests the system may instead be ultracompact, with an orbital period of order 25 minutes.  This is the fourth quiescent black hole candidate discovered to date in a Galactic globular cluster, and the only one with a confirmed accretion signature from its optical/ultraviolet spectrum.
\end{abstract}


\section{Introduction}

Globular clusters (GCs) are extremely efficient at producing low-mass X-ray binaries (LMXBs), with a specific frequency over two orders of magnitude higher than that seen in the field \citep{Cla75}.  Although Galactic GCs comprise less than $10^{-3}$ of the total mass of the Galaxy \citep*{Har13,Lic15}, they contain roughly 10\% of the LMXBs \citep{Ben13}.  The high stellar densities of GCs imply additional formation pathways for LMXBs, via tidal capture and three body exchange interactions.  However, while 18 dynamically-confirmed black holes (BHs) have been identified among the $\sim$200 known LMXBs in our Galaxy \citep[see ][for a review]{McC06}, none of these are in globular clusters.  Either BHs are rare in globular clusters, or selection effects have to date prevented their detection.

\subsection{Stellar-mass black holes in globular clusters}

The most massive stars in a newly-formed GC should collapse into BHs within the first few Myr, and it was originally thought that following mass segregation \citep{Spi69}, mutual gravitational interactions would lead to the expulsion of almost all of these BHs \citep*{Kul93,Sig93,Por00}.  However, recent detections of variable X-ray sources in extragalactic GCs with luminosities in excess of $10^{39}$\,erg\,s$^{-1}$ \citep[e.g.][and references therein]{Mac07,Shi10,Mac11,Bra12,Rob12} have suggested that BHs can in fact survive in GCs.  Closer to home, \citet{Bar14} have recently used X-ray spectral fitting to identify several BH candidates in GCs in M31.

The implied survival of GC BHs has also received theoretical support \citep[e.g.][]{Mac08}, with recent simulations showing that GCs could in fact retain a significant population of BHs \citep{Sip13,Mor13}.  However, this population of predicted BHs is sensitive to the magnitude of the natal kicks received by the BHs at formation.  Assuming that they are not ejected at formation, the BHs will rapidly segregate to the cluster centre, and the exchange of energy between the BH subcluster and the rest of the stars in the cluster allows a significant fraction of the BHs to be retained up to the present day.  \citet{Mor15} performed a set of detailed Monte Carlo simulations, finding that a small fraction ($\sim1$\,\%) of the most massive black holes form a subcluster that undergoes a series of core oscillations, forming a steep cusp and ejecting the most massive single or binary BHs before re-expanding due to the interactions of three-body binaries with other objects.  The rest of the (lower-mass BHs) remain permanently well mixed with the cluster stars.  Depending on initial cluster parameters, up to $10^3$ BHs can remain after a Hubble time.  However, once the BH population has been sufficiently depleted the BHs cannot provide sufficient energy to the rest of the cluster, which then enters a phase of ``core collapse" (defined observationally via a steep radial density profile in the central regions), leading to the rapid ejection of the remaining BHs \citep{Bre13,Heg14}.  This would imply that clusters that have not yet undergone this phase of observational core collapse should be the best candidates to host BHs.

Although 18 LMXBs in GCs have been identified either as persistent sources or during transient outbursts \citep[e.g.][]{Bah14}, sensitive X-ray observations have enabled the detection of dozens of candidate neutron star (NS) LMXBs in quiescence, identified by their soft thermal surface emission \citep[e.g.][]{Rut02,Hei03,Gui09}, enabling the inference that there are of order 10--20 quiescent LMXBs for each LMXB seen in outburst, and thus $\gtrsim200$ quiescent NS LMXBs in GCs \citep{Hei05b}.  Identifying quiescent BH LMXBs in GCs is harder, since they do not produce a distinctive X-ray signature such as the thermal surface emission from quiescent NS LMXBs.  However, accreting BHs do emit compact, partially self-absorbed radio jets in the hard and quiescent states \citep*[e.g.][]{Gal05}, with significantly higher radio luminosities than NS systems with the same X-ray luminosity \citep{Mig06}.  Since the recent bandwidth upgrades to the Very Large Array (VLA) and the Australia Telescope Compact Array (ATCA) have now rendered this faint radio emission accessible over a significant fraction of the Galaxy, we can effectively search for BHs using deep radio continuum observations of GCs \citep{Mac04,Mac05}.  The radio emission should be unresolved, with a flat or slightly inverted spectrum (i.e.\ $\alpha\gtrsim 0$, where we use the convention that flux density $S_{\nu}$ scales with frequency $\nu$ as $S_{\nu}\propto\nu^{\alpha}$).  The inferred radio luminosity should also fall on or within the scatter of the well-established correlation between X-ray and radio luminosities for hard state and quiescent accreting BHs \citep*{Han98,Cor03,Gal03}.  The form of this correlation at the lowest luminosities remains relatively poorly constrained \citep{Gal06}, with recent evidence suggesting more scatter about the correlation than had previously been assumed \citep{Gal14}.  However, in the absence of a dynamical mass, the presence of faint, flat-spectrum radio emission is still the best way to discriminate between an accreting BH and an analogous NS or white dwarf system.  As originally suggested by \citet{MK07}, this technique has recently been used to great effect, discovering two BH candidates in the core of M22 \citep{Str12} and a third candidate in M62 \citep{Cho13}.  While follow-up observations to confirm the nature of these three flat-spectrum radio sources are ongoing, we have continued to obtain deep radio images of GCs to search for further BH candidates.

\subsection{47 Tucanae}
\label{sec:47tuc_intro}

The dense, massive \citep[$6.5\times10^5M_{\sun}$;][]{Kim15} GC 47 Tuc (NGC 104) lies at a distance of $d=4.57\pm0.06$\,kpc \citep[taking the average of multiple previous distance measurements and the standard error on the mean;][]{Woo12}.  Owing to its proximity and low extinction \citep[$E(B-V)=0.04\pm0.02$;][]{Sal07}, the cluster has been very well studied, and is rich in exotic binary systems.  47 Tuc contains 300 known X-ray sources, including five quiescent NS LMXBs \citep{Hei05} and 23 radio millisecond pulsars \citep{Cam00,Fre03}, which are likely LMXB descendants.  Since conditions in the cluster are clearly favourable for forming LMXBs, this makes it an ideal target for deep radio observations to search for quiescent accreting BHs.  Furthermore, \citet{Gie11} recently performed detailed simulations of 47 Tuc, finding that it has not yet undergone core collapse (which will take another $\sim20$\,Gyr) and so should still retain a small population of BHs.  This is in stark contrast to clusters such as M4 and NGC 6397 that have already undergone their ``second core collapse" phase and would no longer be expected to host a significant BH population \citep[see fig.~1 of][]{Heg14}.

The first sensitive radio continuum observations of 47 Tuc were performed by \citet{McC00}, who used 60\,h of ATCA data at 1.4 and 1.7\,GHz to reach rms noise levels of 42 and 46\,$\mu$Jy\,beam$^{-1}$, respectively.  They detected eleven radio point sources within 5\arcmin\ of the cluster centre, including five sources later confirmed as millisecond pulsars (MSPs).  By extending the on-source integration time to 170\,h, \citet*{McC01} reached an rms noise level of 18\,$\mu$Jy\,beam$^{-1}$, detecting an additional seven MSPs and three more unidentified sources, which they suggested to be either extragalactic background sources or as-yet unidentified MSPs.  Within the 24\arcsec\ core radius \citep*{How00}, the only detected sources were MSPs F, G/I, L and O. 

Following the recent bandwidth upgrade to the ATCA \citep{Wil11}, \citet{Lu11} made deep 5.5 and 9.0\,GHz observations of 47 Tuc in 2010 January, with a total on-source integration time of 18.3\,hr, aiming to detect the radiative signature of a possible intermediate-mass black hole (IMBH) at the cluster centre.  After stacking both frequency bands to reach a noise level of 13.3\,$\mu$Jy\,beam$^{-1}$, they detected no source brighter than $3\sigma$ within the central 10\arcsec$\times$10\arcsec\ of the cluster, giving a $3\sigma$ upper limit on the mass of any IMBH of $520$--$4900M_{\sun}$.  A similar IMBH limit was found by \citet{McL06}, who used the cluster velocity dispersion profile to place a $1\sigma$ upper limit of $1500 M_{\sun}$ on the mass of any IMBH, so the evidence for such an object is not compelling.

In this paper, we combine the archival observations of \citet{Lu11} with new ATCA data to make the deepest radio image of 47 Tuc to date.  We report the discovery of a flat-spectrum radio source within the cluster core that is an astrometric match to the known bright X-ray source X9 \citep*{Her83,Aur89}.  In light of its inferred radio luminosity, we suggest that the source is a quiescent BH, rather than a cataclysmic variable (CV), as had been previously inferred \citep*{Par92,Gri01,Kni08}.

In Sections \ref{sec:obs} and \ref{sec:results}, we describe our radio observations and results.  We discuss previous observations of X9 in Section \ref{sec:x9}, comparing the source properties to those of various classes of accreting compact object.  We discuss the possible nature of the system in Section \ref{sec:nature}, review the sample of GCs hosting BH candidates in Section \ref{sec:discussion}, and present our conclusions in Section \ref{sec:conclusions}.

\section{Observations and data reduction}
\label{sec:obs}

\subsection{ATCA data}
\label{sec:atca}
We observed the globular cluster 47 Tuc with the ATCA on 2013 November 12th, from 08:06--17:52 UT (MJD $56608.54\pm0.20$), achieving an on-source integration time of 8.7\,hr.  Using the Compact Array Broadband Backend (CABB), we observed simultaneously in two bands, with central frequencies of 5.5 and 9.0 GHz, each with a bandwidth of 2048\,MHz.  The array was in its extended 6A configuration, with a maximum baseline of 5.939\,km.  We used B1934-638 as both a bandpass calibrator and to set the flux density scale, and B2353-686 as the secondary calibrator to set the amplitude and phase gains.

We reduced the data for each frequency band separately, performing external gain calibration in Miriad \citep*{Sau95} using standard procedures.  We then frequency-averaged the calibrated data and imported them into the Common Astronomy Software Application \citep[CASA;][]{McM07} for imaging.  Imaging was carried out using Briggs weighting with a robust parameter of 1, which provided a good compromise between sensitivity and resolution, as well as suppressing the sidelobes of the dirty beam.  Our final image sensitivities of 5.7 and 6.4\,$\mu$Jy\,beam$^{-1}$ at 5.5 and 9.0\,GHz, respectively, were close to the theoretical thermal noise levels.  The final source fluxes and positions were measured in the image plane using the \textsc{jmfit} algorithm within the Astronomical Image Processing System \citep[AIPS;][]{Gre03}.

\subsection{Archival ATCA data}

To improve the significance of our detections, we combined our data with the archival ATCA data taken in the same array configuration by \citet{Lu11} on 2010 January 24--25 (summarised briefly in Section~\ref{sec:47tuc_intro}).  Our re-analysis of these data allowed us to reduce the noise levels in the 5.5 and 9.0\,GHz images to the theoretically-expected limits of 6.7 and 9.8\,$\mu$Jy\,beam$^{-1}$, respectively.  Despite the longer on-source integration times, these observations were less sensitive than those from 2013 (Section~\ref{sec:atca}) because the older data were taken prior to the installation of the new, more sensitive 4-cm receivers at the ATCA.

\section{Results}
\label{sec:results}

\begin{figure}
\centering
\centerline{\includegraphics[width=\columnwidth]{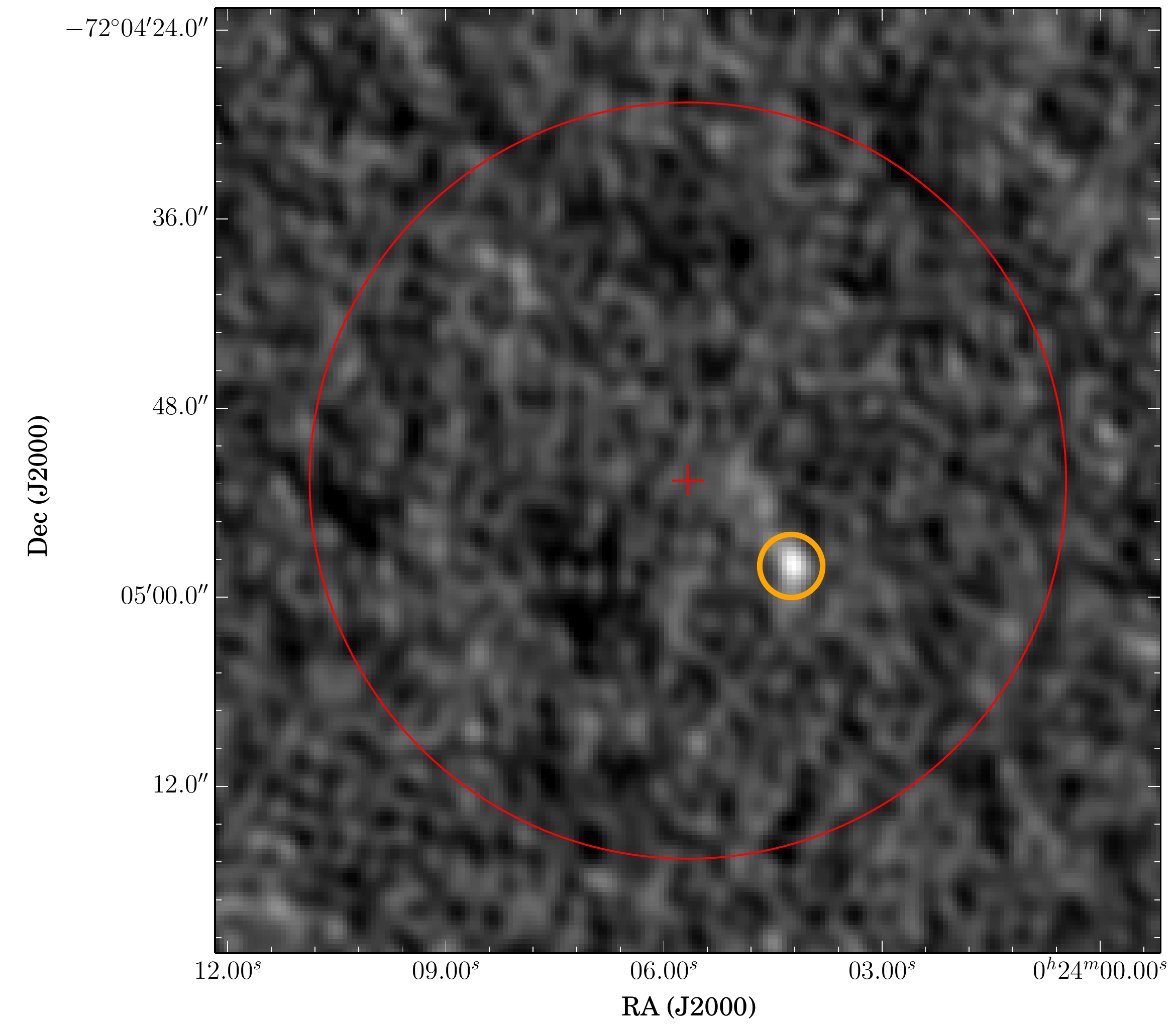}}
\caption{5.5-GHz ATCA image of the core of 47 Tuc.  The red cross denotes the cluster centre, the red circle shows the $24^{\prime\prime}$ core radius, and the thick orange circle highlights our detection of X9.}
\label{fig:core}
\end{figure}

After stacking both the new and archival ATCA data in the $uv$-plane, our resulting images reached rms noise levels of 4.4\,$\mu$Jy\,beam$^{-1}$ at 5.5\,GHz and 5.7\,$\mu$Jy\,beam$^{-1}$ at 9.0\,GHz.  The final images at 5.5 and 9.0\,GHz had resolutions of $2\farcs3\times1\farcs7$ in P.A.\ 18\degree, and $1\farcs5\times1\farcs1$ in P.A.\ 28\degree, respectively.

\subsection{A radio counterpart to X9/W42}
\label{sec:counterpart}

\begin{figure}
\centering
\centerline{\includegraphics[width=\columnwidth]{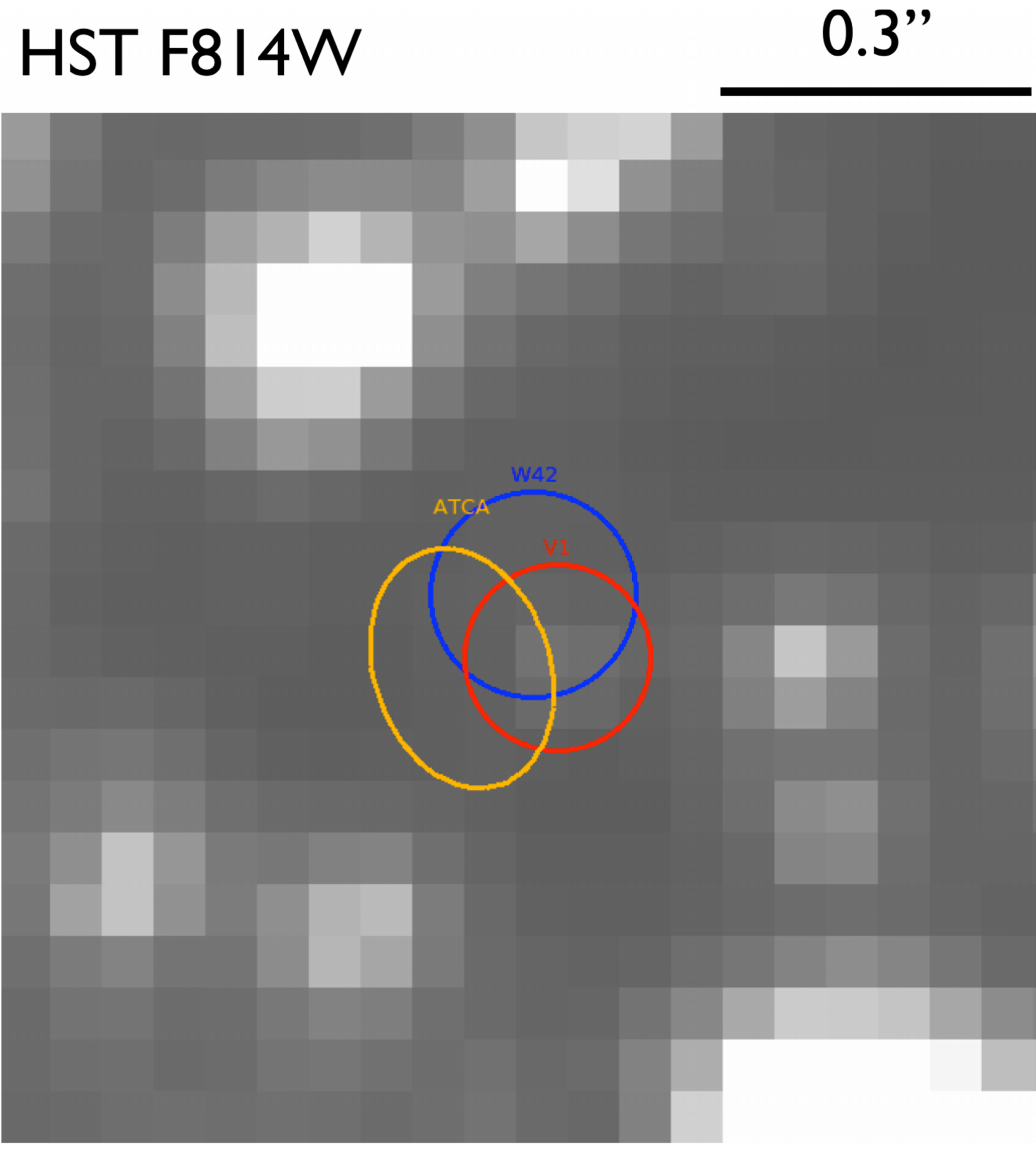}}
\caption{{\it HST} F814W image, showing the $1\sigma$ error ellipses of 47 Tuc X9/W42/V1 in the radio (orange), optical (red) and X-ray (blue) bands. North is up and east is to the left.  Uncertainty on the radio position arises purely from the statistical error in fitting the source position.  Optical and X-ray error ellipses are dominated by the uncertainty in the absolute astrometry (matching the {\it HST} reference frame to 2MASS in the optical band, and using the measured positions of X-ray counterparts to known radio MSPs to estimate the uncertainty on the X-ray reference frame). The error ellipses in all three wavebands overlap.}
\label{fig:astrometry}
\end{figure}

The brightest radio source within the 24\arcsec\ core radius of the cluster at both frequencies (Fig.~\ref{fig:core}) is coincident with the second brightest hard X-ray source in the cluster (see Fig.~\ref{fig:astrometry}), denoted as source X9 in the {\it ROSAT} catalogue of \citet*{Has94}, and W42 in the {\it Chandra} catalogues of \citet{Gri01} and \citet{Hei05}.  We summarise in Table~\ref{tab:x9} the measured radio brightness of the source in each observation.  Using the stacked image, a point source fit to the radio source position in the image plane using the {\sc AIPS} task {\sc jmfit} gave a J2000 position of
\begin{align}
{\rm R.A.}&=00^{\rm h}24^{\rm m} 04\fs264\pm0\fs016\nonumber\\
{\rm Dec.}&= -72^{\circ}04^{\rm m}58\farcs09\pm0\farcs10.
\nonumber
\end{align}

\begin{table}
{\caption{\label{tab:x9}Measured flux densities of 47 Tuc X9.  The source was fit as a point source in the image plane, using the {\sc AIPS} task {\sc jmfit}.}}
\begin{tabular}{cccc}
\hline
Data & 5.5-GHz & 9.0-GHz & Spectral \\
set & flux density & flux density & index \\
(J2000) & ($\mu$Jy\,bm$^{-1}$) & ($\mu$Jy\,bm$^{-1}$) & \\
\hline
2010 January 24--25 & $34\pm7$ & $36\pm10$ & $0.2\pm0.7$\\
2013 November 12 & $50\pm6$ & $36\pm6$ & $-0.6\pm0.5$\\
Stacked data & $42\pm4$ & $35\pm6$ & $-0.4\pm0.4$\\
\hline
\end{tabular}
\end{table}

Our fitted peak flux densities for X9 in the stacked data were $42\pm4$\,$\mu$Jy\,beam$^{-1}$ at 5.5\,GHz and $35\pm6$\,$\mu$Jy\,beam$^{-1}$ at 9.0\,GHz.  We therefore derive a radio spectral index of $\alpha=-0.4\pm0.4$ and a 5.5-GHz radio luminosity of $5.8\times10^{27}$\,erg\,s$^{-1}$.  Here, and throughout this paper, we define radio luminosity by assuming a flat radio spectrum up to the observing frequency (i.e.\ $L_{\rm r} = 4\pi d^2 \nu S_{\nu}$, where $d$ is the source distance, $\nu$ is the observing frequency, and $S_{\nu}$ is the observed flux density).  No circular polarisation was detected from the source in either of the individual epochs, with the best $5\sigma$ limit in the stacked data set being $<22$\,$\mu$Jy\,beam$^{-1}$ ($<52$\%) at 5.5\,GHz.  We also placed a $5\sigma$ limit of $<30$\,$\mu$Jy\,beam$^{-1}$ ($<71$\%) on the 5.5-GHz linear polarisation.

To quantify the variability of the radio counterpart to X9, we examined the individual images from the 2010 and 2013 data sets.  Point source fits to the position of X9 showed the 9-GHz flux to be constant within uncertainties, while the 5.5-GHz flux increased from $34\pm7$ to $50\pm6$\,$\mu$Jy\,beam$^{-1}$, implying marginal ($1.6\sigma$) evidence for variability and a slight steepening of the spectrum between the two epochs (see Table~\ref{tab:x9}).  We also split the more sensitive 2013 data set into two halves, finding no evidence for significant variability on timescales of a few hours.

\subsection{Other radio sources}

\begin{figure*}
\centering
\centerline{\includegraphics[width=\textwidth]{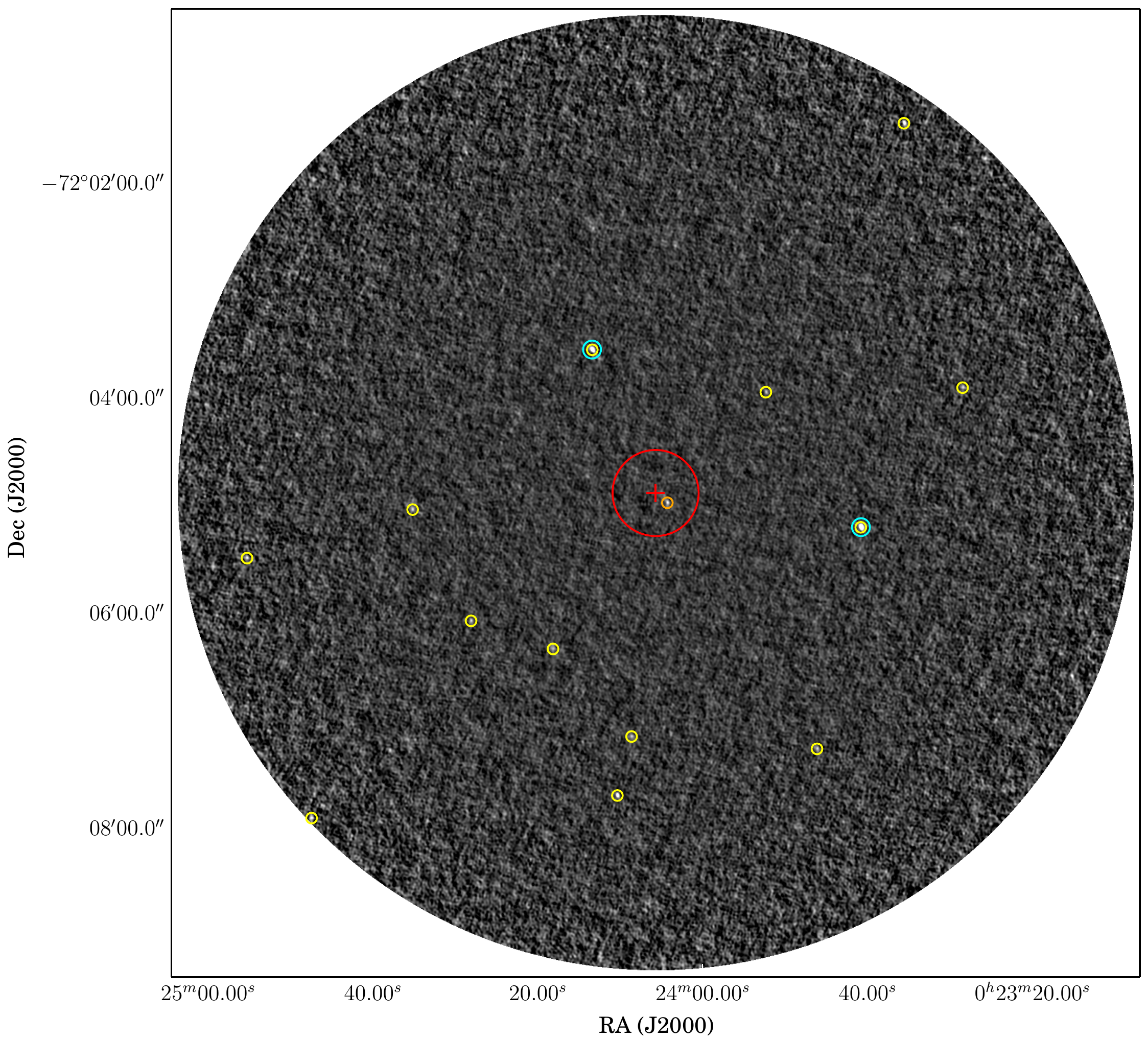}}
\caption{5.5-GHz ATCA image of the field surrounding 47 Tuc, encompassing the 8\farcm9 full width at half power (FWHP) of the primary beam at 5.5\,GHz (at 9.0\,GHz, the FWHP is 5\farcm3). Red circle shows the 24\arcsec\ core radius, with the cluster centre marked by the red cross.  Sources detected by {\sc aegean} at 5.5 (9.0) GHz are denoted by yellow (cyan) circles, except for X9, which is the one orange circle inside the core radius.}
\label{fig:field}
\end{figure*}

While X9 was the only source in the cluster core detected at $>5\sigma$, there were several other radio sources in the field.  We used the \textsc{aegean} source finder \citep{Han12,Han12src} to detect and characterise all sources within the half-power response of the primary beam at both frequencies (radii of 4\farcm5 and 2\farcm7 at 5.5 and 9.0\,GHz, respectively).  \textsc{aegean} fits elliptical Gaussian components (not constrained to the synthesised beam size, hence not necessarily point sources) to all islands of pixels with a peak exceeding a user-set threshold relative to the local background rms noise level, returning the amplitude of an elliptical Gaussian fit to each component.  With over $10^7$ pixels in our images, we set a detection threshold of $5.5\sigma$.  Including X9, \textsc{aegean} detected fourteen radio sources exceeding this threshold within the central 4\farcm5 FWHM of the 5.5-GHz image (Fig.~\ref{fig:field}).  At 9.0\,GHz, only three sources (all with 5.5-GHz counterparts) were detected within the 2\farcm7 diameter FWHM.  Details of all detected sources are given in Table~\ref{tab:radio_sources}.

We also examined the positions of the known MSPs in our 5.5-GHz image \citep[identified from the ATNF pulsar catalogue\footnote{http://www.atnf.csiro.au/research/pulsar/psrcat};][]{Man05}, and although none were detected at the $5.5\sigma$ level (and hence picked up by \textsc{aegean}), seven showed radio emission at $\gtrsim3\sigma$ (Table~\ref{tab:msps}).  We carried out a point source fit to each, using the \textsc{jmfit} task within \textsc{AIPS}.   For all other MSPs in the cluster, we place a $3\sigma$ upper limit of 15\,$\mu$Jy\,beam$^{-1}$ at 5.5\,GHz.  No MSPs were detected in the 9.0-GHz image.

Of the 14 sources detected at 5.5\,GHz by \textsc{aegean}, four had an X-ray counterpart in the {\it Chandra} catalogue of \citet{Gri01}.  These were W42 (X9), W132, W186 and W231, but only X9 had previously been identified with any particular source class. Given the flat radial distribution of X-ray sources beyond 100\arcsec\ (where the other three radio sources with X-ray counterparts are located), \citet{Hei05} inferred that such distant sources are likely not associated with the cluster, but are more likely background AGN. Indeed, optical imaging with the {\it Hubble Space Telescope} ({\it HST}) shows that W186 and W231 are likely associated with extragalactic sources since they since they lie in a region of extended optical emission (van den Berg et al. in prep).  Although W132 is outside the field of view of that data set, it is a very hard X-ray source \citep{Hei05}, suggestive of a background AGN with intrinsic absorption \citep{Gri02,Hei05}.  Should these identifications as background AGN prove correct, these sources could be used as calibrators to define a stationary reference frame for astrometric studies, allowing an improved distance determination for the cluster.  

We compared the positions of these three X-ray sources (W132, W186 and W231) from the {\it Chandra} catalogue of \citet{Hei05} with those of their ATCA radio counterparts (Table~\ref{tab:radio_sources}), and found that the radio and X-ray frames could be matched to a $3\sigma$ limit of better than 0\farcs2--0\farcs3.  A comparison of our radio position for X9 with the X-ray position measured by {\it Chandra} in 2002 \citep{Hei05} suggests that the two are offset by $0\farcs11\pm0\farcs11$ in R.A.\ and $-0\farcs12\pm0\farcs17$ in Dec., i.e.\ consistent to within $1\sigma$ (see Fig.~\ref{fig:astrometry}).

\begin{table*}
{\caption{\label{tab:radio_sources}Fitted positions and flux densities of the radio sources detected at $>5.5\sigma$ by \textsc{aegean}.  Quoted uncertainties are the $1\sigma$ uncertainties on the elliptical Gaussian fit parameters.  Identifications show the associated X-ray wavdetect identifications (W-labels) from \citet{Gri01} and \citet{Hei05}, and the radio source numbers (MA-labels) from \citet{McC00}.  Upper limits are given at the $5.5\sigma$ level.  The upper limits at 9\,GHz vary depending on the primary beam response at the source position.}}
\begin{tabular}{cccccc}
\hline
RA & Dec & 5.5-GHz peak & 9.0-GHz peak & ID & Offset from \\
& & flux density & flux density & & cluster centre\\
(J2000) & (J2000) & ($\mu$Jy\,bm$^{-1}$) & ($\mu$Jy\,bm$^{-1}$) & & (arcmin)\\
\hline
00 23 28.608 $\pm$ 0.008 & -72 03 53.62 $\pm$ 0.16 & $ \phantom{0} 32\pm  4$ & $<76$ & & 3.01 \\
00 23 35.756 $\pm$ 0.007 & -72 01 26.24 $\pm$ 0.13 & $ \phantom{0} 53\pm  5$ & $<191$ & MA2 & 4.14 \\
00 23 40.868 $\pm$ 0.002 & -72 05 11.62 $\pm$ 0.04 & $116\pm  3$ & $ 52\pm  2$ & MA3 & 1.93 \\
00 23 46.130 $\pm$ 0.008 & -72 07 15.21 $\pm$ 0.16 & $ \phantom{0} 31\pm  4$ &  $<66$ & & 2.81 \\ 
00 23 52.358 $\pm$ 0.008 & -72 03 56.36 $\pm$ 0.15 & $ \phantom{0} 27\pm  3$ & $<37$ & & 1.39 \\ 
00 24 04.247 $\pm$ 0.006 & -72 04 58.09 $\pm$ 0.12 & $ \phantom{0} 34\pm  3$ & $ 30\pm  2$ & W42 (X9) & 0.14 \\
00 24 08.583 $\pm$ 0.009 & -72 07 08.40 $\pm$ 0.15 & $ \phantom{0} 29\pm  3$ & $<53$ & W186 & 2.27\\ 
00 24 10.304 $\pm$ 0.004 & -72 07 41.30 $\pm$ 0.08 & $ \phantom{0} 57\pm  3$ & $<66$ & MA6 & 2.83\\ 
00 24 13.313 $\pm$ 0.003 & -72 03 32.73 $\pm$ 0.05 & $ \phantom{0} 71\pm  3$ & $ 35\pm  1$ & MA7 & 1.46 \\
00 24 18.081 $\pm$ 0.009 & -72 06 19.54 $\pm$ 0.13 & $ \phantom{0} 26\pm  3$ & $<44$ & W231 & 1.73\\ 
00 24 27.990 $\pm$ 0.007 & -72 06 03.82 $\pm$ 0.14 & $ \phantom{0} 30\pm  3$ & $<48$ & & 2.09\\ 
00 24 35.034 $\pm$ 0.008 & -72 05 01.71 $\pm$ 0.15 & $ \phantom{0} 33\pm  4$ & $<48$ & W132 & 2.26\\ 
00 24 47.350 $\pm$ 0.010 & -72 07 53.63 $\pm$ 0.12 & $ \phantom{0} 40\pm  4$ & $<249$ & & 4.40\\
00 24 55.057 $\pm$ 0.010 & -72 05 28.50 $\pm$ 0.14 & $ \phantom{0} 37\pm  4$ & $<144$ & & 3.84\\
\hline
\end{tabular}
\end{table*}

\begin{table*}
{\caption{\label{tab:msps}Fitted 5.5-GHz positions and flux densities of the MSPs detected at $>3\sigma$.  All known MSP positions were inspected by eye.  Where radio emission was detected at the pulsar position, a point source fit was carried out using the \textsc{aips} task \textsc{jmfit}.}}
\begin{tabular}{ccccc}
\hline
RA & Dec & Peak flux density & ID \\
(J2000) & (J2000) & ($\mu$Jy\,bm$^{-1}$) & \\
\hline
00 23 59.49 $\pm$ 0.04 & -72 03 58.8 $\pm$ 0.2 & $19\pm5$ & MSP-J\\
00 24 04.68 $\pm$ 0.05 & -72 04 54.4 $\pm$ 0.3 & $16\pm5$ & MSP-O\\
00 24 06.71 $\pm$ 0.07 & -72 04 06.7 $\pm$ 0.4 & $11\pm5$ & MSP-H\\
00 24 08.02 $\pm$ 0.05 & -72 04 39.6 $\pm$ 0.3 & $15\pm5$ & MSP-G/I\\
00 24 08.47 $\pm$ 0.05 & -72 04 38.6 $\pm$ 0.3 & $14\pm5$ & MSP-T\\
00 24 13.92 $\pm$ 0.03 & -72 04 43.4 $\pm$ 0.2 & $23\pm5$ & MSP-D\\
00 24 16.63 $\pm$ 0.05 & -72 04 25.0 $\pm$ 0.3 & $16\pm5$ & MSP-Q\\
\hline
\end{tabular}
\end{table*}

\section{47 Tuc X9}
\label{sec:x9}

An X-ray source was first detected in the core of 47 Tuc by the {\it Einstein} satellite, with a variable X-ray luminosity that reached $\sim1.0 \times 10^{34}$\,erg\,s$^{-1}$ \citep[0.5--10\,keV;][]{Her83,Aur89}.  A faint, variable and very blue optical candidate counterpart, denoted V1, was identified by \citet{Par92}.  {\it ROSAT} observations resolved the cluster into multiple X-ray sources, of which one, X9, was coincident with V1 \citep{Has94,Ver98}.  \citet{Gri01} used {\it Chandra} to confirm the association of this source (which they label W42) with V1.  A far-ultraviolet (UV) observation by \citet{Kni02} identified V1 as far-UV bright and variable.

\subsection{Optical/Ultraviolet/X-ray properties}
\label{sec:properties}

\citet{Par92} suggested that the broadband optical through ultraviolet spectrum of the source was consistent with an accretion disc in a quiescent CV, and on the basis of its high X-ray to optical flux ratio, suggested that the system could be an intermediate polar (a magnetic CV in which the accretion disc is disrupted inside the white dwarf magnetosphere, causing the infalling matter to accrete along the magnetic field lines).  V1 is located blueward of the cluster main sequence in both $U$ against ($U-V$) and $V$ against ($V-I$) colour-magnitude diagrams \citep{Edm03a} and shows significant optical flickering \citep{Edm03b}, demonstrating that there is a substantial contribution to the optical light from the accretion disc.  The (non-simultaneous) broadband infrared through far-ultraviolet spectral energy distribution (SED) could be qualitatively fit by an optically-thick model atmosphere of temperature 12,000\,K and effective radius $0.18R_{\sun}$, presumably associated with the accretion disc \citep{Kni08}.

Time-series analysis of the optical counterpart by \citet{Edm03b} found a marginally-significant 3.5-hour sinusoidal signal in the $V$-band power spectrum that implied an orbital period of $\sim3.5$ or 7\,hr, depending on whether or not the periodicity could be ascribed to ellipsoidal modulations. For a Roche-lobe filling main-sequence donor star, \citet{Kni08} found that a 7-hr period would imply a donor that was too bright at wavelengths $>5000$\,\AA.  For non-ellipsoidal variability, a 3.5-hr period would then imply a donor of mass $0.5M_{\sun}$, which, while not required, could be accommodated by their model for the broadband spectrum, which was dominated by the hot accretion disc.  In slitless ultraviolet spectroscopy with {\it HST}, \citet{Kni08} also detected the spectroscopic signature of a relatively high-inclination accretion disc or disc wind, namely a strong, double-peaked C\,{\sc iv} emission line, which they took to confirm the identification of the source as a CV.

With a 0.5--6\,keV luminosity of $0.6$--$1.0\times10^{33}$\,erg\,s$^{-1}$\citep[][after scaling to our adopted distance of 4.57\,kpc]{Hei05}, the high X-ray--to--optical flux ratio ($F_{\rm X}/F_{\rm opt}\sim 5$) of V1 is consistent with the system being a CV, but is higher than any known field CV, and comparable only to another (confirmed magnetic) CV in 47 Tuc, V3/X10 \citep{Edm03b,Hei05}.  While \citet{Gri01} reported a marginal detection of a 218-s periodicity in the X-ray emission from X9, the original claim was not highly significant, and attempts with larger datasets to recover this periodicity have failed (Bahramian et al.\ in prep.).

The 2000 {\it Chandra} X-ray spectrum of X9 could be fit by a thermal bremsstrahlung model with a temperature of $32\pm6$\,keV \citep{Gri01}.  However, a higher-quality 2002 {\it Chandra} spectrum revealed additional features \citep{Hei05}.  A strong emission line (likely O\,{\sc viii}) at 0.65\,keV superimposed on a hard continuum could be fit with a thermal plasma model, with two MEKAL components at 0.25 and $>17$\,keV, together with a third, highly-absorbed component needed to fit the spectrum above 6\,keV.  This is not typical of a CV, but equally is not obviously indicative of any other class of source.  However, the long exposure time ($\sim70$\,ks in 2000 and $\sim280$\,ks in 2002) and the low absorption towards 47 Tuc \citep[$N_{\rm H}=1.3\times10^{20}$\,cm$^{-2}$;][]{Hei05} imply that the spectrum is of higher quality than typically available for accreting stellar-mass compact objects.  Since fits to MEKAL components with high temperatures are typically indistinguishable from power-law spectral fits, we therefore re-fit a short, 5.3-ks observation (ObsID 3384, from 2002 September 30, taken in subarray mode to alleviate pileup) with a model consisting of an absorbed power-law plus a Gaussian line to represent the low-energy MEKAL component, to enable a comparison with typical spectra from other classes of source (see Sections~\ref{sec:tmsp} and \ref{sec:bh}).  Fixing the absorption to the cluster value, we found a fitted photon index of $\Gamma=1.06\pm0.24$, a line energy of $0.56\pm0.03$\,keV, and a line width of $0.10\pm0.03$\,keV.

\subsection{Radio emission from cataclysmic variables}
\label{sec:cvs}

Our radio detection of X9 provides new constraints on the nature of the accreting object.  Since the source was previously identified as a CV, we first compare its radio properties with those of known CVs, of which only a handful have been detected in the radio band to date.

Of the magnetic CVs, the most luminous persistent radio emitter is the rapidly-rotating intermediate polar AE Aqr, with a typical 5-GHz flux density of 4--8\,mJy\,beam$^{-1}$.  Its spin period of just 33\,s \citep{Pat79} creates a propeller effect \citep*{Wyn95,Era96}, with the expulsion of magnetised plasma clouds giving rise to the observed, rapidly-variable synchrotron emission \citep{Mei05}, which has a typical spectral index of 0.35--0.60.  At a distance of 86\,pc, its maximum measured 5-GHz flux density of 16\,mJy \citep*[during a short-duration, $\sim 2$-hour flare; see][]{Bas88,Aba93} would correspond to a radio luminosity of $7\times10^{26}$\,erg\,s$^{-1}$.

Other than AE Aqr, only the two polars AM Her \citep{Cha82} and AR UMa \citep{Mas07} have been found to be persistent radio emitters among the magnetic CVs.  Their quiescent emission ($L_{\rm r}\sim2\times10^{25}$\,erg\,s$^{-1}$) has been attributed to gyrosynchrotron emission from energetic electrons in the white dwarf magnetosphere, with occasional highly circularly-polarised flaring events due to cyclotron maser emission \citep{Mas07}.  The quiescent emission of AM Her appeared to be relatively stable over a ten-year period \citep*{Cha82,Pav94}, and from observations at 4.9 and 8.4\,GHz taken 22 years apart, we derive a non-simultaneous spectral index of $-0.3\pm0.3$.

The most luminous reported radio emission from a CV was a decaying flare detected from the intermediate polar DQ Her by \citet{Pav94}.  Over 30\,min, the flux density decreased by a factor of 4, with an initial radio luminosity of $L_{\rm r}=5\times10^{27}$\,erg\,s$^{-1}$.  The system was not detected on three further occasions.  Similarly short-timescale flares with comparable luminosities have also been reported from V834 Cen \citep{Wri88} and BG CMi \citep{Pav94}, although these detections were both lower-significance, and the former was from a single-dish observation and therefore potentially more susceptible to the effects of both source confusion and radio frequency interference.

Of the non-magnetic CVs, the only published radio detections to date are from the dwarf novae SS Cyg, EM Cyg and SU UMa, and the nova-like systems V3885 Sgr, TT Ari, RW Sex and V603 Aql.  Of these, the most well-studied is SS Cyg, whose radio counterpart has been repeatedly detected at the peak of its frequent outbursts, and whose high brightness temperature radio emission has been interpreted as synchrotron emission from a jet \citep{Koe08}.  At a distance of $114\pm2$\,pc, its brightest detected radio flare of 3\,mJy\,beam$^{-1}$ \citep{Mil13} corresponds to a radio luminosity of $2.3\times10^{26}$\,erg\,s$^{-1}$.  Single detections of both SU UMa \citep*{Ben83} and EM Cyg \citep{Ben89} implied similar radio luminosities of $\sim 1$--$5\times10^{26}$\,erg\,s$^{-1}$ (depending on the poorly-constrained source distances).  The four nova-like systems were detected at levels of 0.03--0.2\,mJy\,beam$^{-1}$ \citep{Koe11,Cop15}, which for the estimated distance ranges imply radio luminosities of $4.5\times10^{24}$--$1.6\times10^{26}$\,erg\,s$^{-1}$.

In summary, our radio detection of X9 is an order of magnitude more luminous than the persistent emission detected from any known CV.  While it is comparable to the peak luminosities of infrequent short-timescale flaring events, the stability of the source over the course of each 10-h observing run, and between the two epochs in 2010 and 2013, argues against an origin as flaring events.  Furthermore, the absence of a high level of circular polarisation suggests that the emission is unlikely to arise from an electron cyclotron maser.  Thus, in light of our radio data, we do not favour an interpretation of X9 as a CV.

\subsection{A transitional millisecond pulsar?}
\label{sec:tmsp}

\begin{figure*}
\centering
\includegraphics[width=0.8\textwidth]{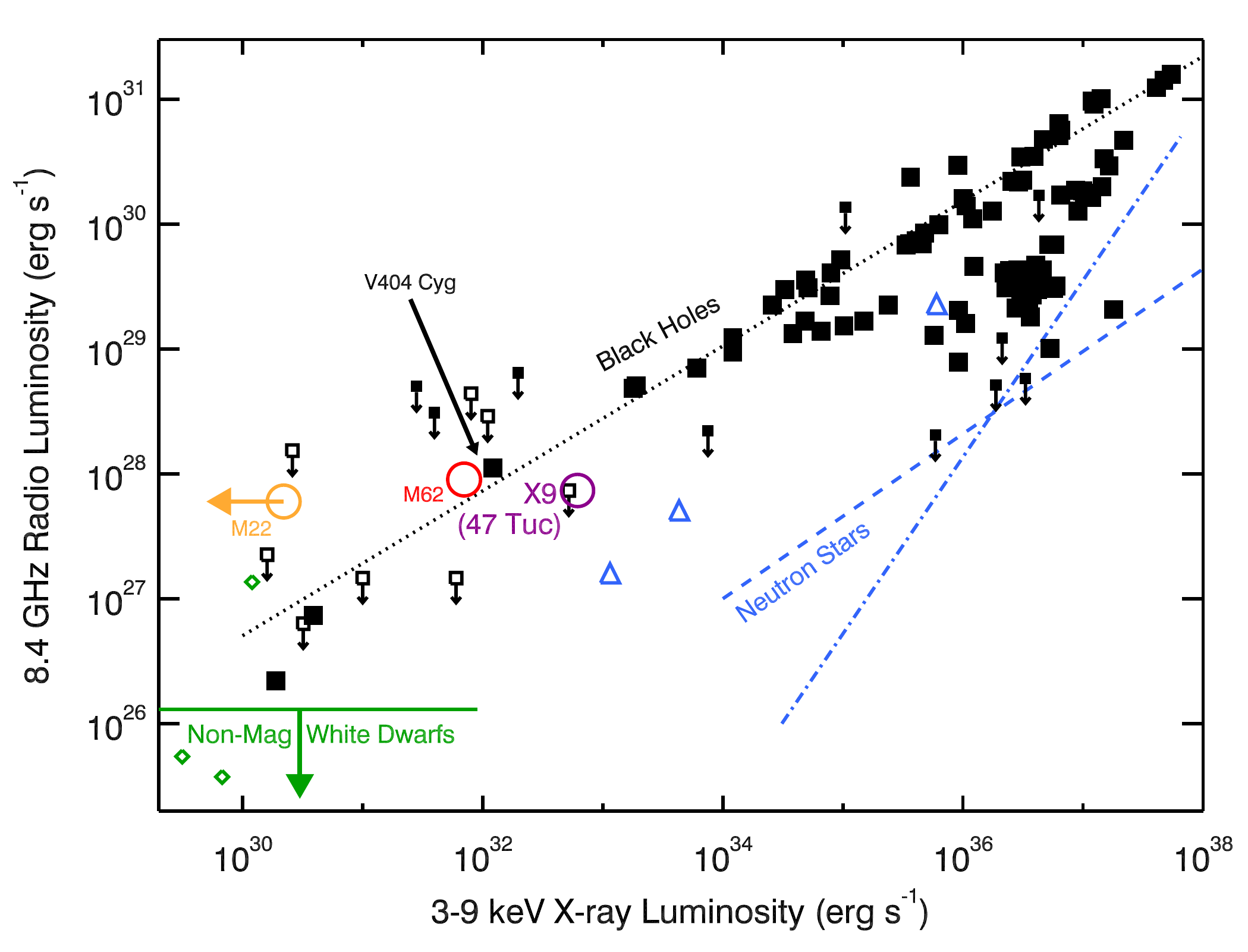}
\caption{Radio/X-ray correlation in hard state and quiescent stellar-mass black holes.  Black squares show field black holes from the literature \citep*{Mil11,Gal12,Rat12,Cor13,Gal14}, blue triangles show the transitional binary MSP systems \citep{Hil11,Pap13,Del15}, green diamonds show the magnetic CVs \citep{Aba93,Cha82,Mas07}, yellow circle shows the M22 candidates \citep{Str12}, red circle shows the M62 candidate \citep{Cho13}, and purple circle shows 47 Tuc X9.  Filled symbols are from simultaneous radio and X-ray data, whereas open symbols are non-simultaneous measurements.  Black dotted line shows the best fit relation for the black hole systems \citep[$L_{\rm r}\propto L_{\rm X}^{0.58}$;][]{Gal06}.  Blue dashed and dash-dotted lines show two possible radio/X-ray correlations for neutron star systems \citep{Mig06}, and green line shows the maximum radio luminosity of non-magnetic accreting white dwarf systems in outburst \citep{Koe08,Koe11}.  X9 falls within the scatter of the best fitting radio/X-ray correlation for black holes, and is too radio luminous to be a CV.}
\label{fig:lrlx}
\end{figure*}

Another class of accreting sources known to produce flat-spectrum radio emission are the binary MSPs that show occasional transitions to and from an accretion-powered state characteristic of an LMXB.  To date, we know of three such confirmed transitional binary systems, which can spend years in a phase of quiescent accretion, with X-ray luminosities in the range $10^{32}$--$10^{34}$\,erg\,s$^{-1}$ \citep{deMar13,Pat14}. Both in this quiescent phase and also during the brighter outburst phase the observed millisecond X-ray pulsations at the pulsar spin period show that some fraction of the accreted material is reaching the neutron star surface \citep{Pap13,Arc15,Pap15}.  All three systems have also shown flat-spectrum continuum radio emission when in their accretion-powered states \citep{Bon02,Pap13,Hil11,Bas14,Del15}, with inferred radio luminosities significantly higher than expected from an extrapolation of the radio/X-ray correlation for the traditional atoll source neutron stars \citep[Fig.~\ref{fig:lrlx}; see][]{Del15}.  However, the proposed correlation between their measured radio and X-ray luminosities \citep{Del15} implies radio emission a factor of $\sim6$ lower than seen for X9 at a similar X-ray luminosity (see Fig.~\ref{fig:lrlx}).

The three transitional MSP systems share some unusual X-ray variability characteristics, which can be compared to the X-ray variations of X9.  Each transitional system has switched at least once between a radio pulsar state (with $L_X\sim10^{32}$\,erg\,s$^{-1}$) and a low-luminosity accretion state (with $L_X$ averaging $\sim4\times10^{33}$\,erg\,s$^{-1}$), with M28I also showing a bright accretion outburst at $L_X > 10^{36}$\,erg\,s$^{-1}$.  Within the low-luminosity accretion state, each system switches among three modes; a ``high'' mode showing pulsations, sudden dips to a ``low'' mode (transitions less than 30 s, average flux a factor 5--10 lower), and a rarer ``flaring'' mode (lasting 1--45 minutes, typically a factor $>$4 brighter).  This state also shows a remarkably constant average X-ray flux (set by the ``high'' mode flux); multiple observations over years have not found factor-of-two variations \citep{deMar13,Lin14,Bog15}.  This is in contrast to X9, where the average flux varies smoothly by a factor of ten over timescales of several days (Bahramian et al. 2015, in prep.).

Each transitional MSP system also shows all three modes.  Dips to the ``low'' mode can be quite short ($<100$\,s), below our capacity to detect in the lightcurve of X9.  However, dips longer than 100\,s occur frequently; in PSR J1023+0038, dips longer than 200\,s occur every 3\,ks \citep{Bog15}, and in XSS J12270$-$4859, dips between 100 and 1100\,s also occur on average every 3\,ks \citep{deMar13}.  The M28I lightcurve of \citet{Lin14}, limited to binning of 500\,s, does not probe such short timescales, but nevertheless showed six low states of length exceeding 1\,ks, spread over 234\,ks.  Flares are also common, occurring every 20\,ks or so in PSR J1023+0038 \citep{Bog15}, and at similar frequency (occupying 1/10 to 1/3 of the time) in XSS J12270$-$4859 \citep{deMar13}.  Careful inspection of the 300\,ks of ACIS-S lightcurves of X9 from 2002 \citep{Hei05}, binned at 50\,s, fails to show evidence of either sharp transitions to low modes, or flaring by a factor $>$2.  Extrapolation from the transitional MSPs suggests that of order 100 low modes, and 15 flaring episodes, should have been observed; the lack of these behaviours strongly suggests that X9 is not a similar system as the transitional MSPs.

Furthermore, the optical spectrum of PSR J1023+0038 during its accretion-powered state shows multiple double-peaked He emission lines \citep{Wan09}, in contrast to X9, which shows no strong evidence for He emission in its far-ultraviolet spectrum \citep{Kni08}.  The X-ray spectrum of X9 (Section~\ref{sec:properties}) is also markedly different from the spectra of the transitional MSPs when at comparable luminosities in the accretion-powered state.  With a similarly low absorption column density of $N_{\rm H}\sim3\times10^{20}$\,cm$^{-2}$, PSR J1023+0038 shows no evidence for an emission line near 0.6\,keV, and the spectrum can be well fit by a simple power-law model with no requirement for an additional MEKAL component \citep{Bog15}.  And despite the larger column densities towards XSS J12270$-$4859 and M28I, there is no evidence for such an emission line in high-quality spectra of either source \citep{deMar10,Pap13}.  Furthermore, at X-ray luminosities of $10^{33}$--$10^{34}$\,erg\,s$^{-1}$ all three transitional MSPs can be fit by simple power-law spectra with photon indices of $\Gamma=$1.5--1.7 \citep[][]{deMar10,Pap13,Bog15}, significantly softer than seen in X9 (Section~\ref{sec:properties}).

In summary, while we cannot explicitly rule out X9 being a transitional MSP, the available X-ray, optical and ultraviolet data, taken together with the relatively high radio luminosity, lead us to consider this explanation to be unlikely.

\subsection{X9 as a black hole}
\label{sec:bh}

The high radio luminosity implied by our observations of X9 suggests an alternative identification of X9, as a stellar-mass BH.   Comparison of our radio measurements with the X-ray luminosity measured by \citet{Hei05} shows that the source lies very close to the well-established radio/X-ray correlation for accreting stellar-mass BHs (Fig.~\ref{fig:lrlx}).  And while the X-ray and radio observations were separated by 8--10 years, the relative stability of the radio flux density between 2010 and 2013, and the X-ray flux between 1992 and 2002 \citep{Ver98,Gri01,Hei05} helps assuage (but does not eliminate) concerns about strong source variability.

\citet{Kni08} found the broadband optical-through-ultraviolet spectrum of V1 to be dominated by an optically-thick component of temperature $\sim$12,000\,K and radius $\sim0.18R_{\sun}$, consistent with a disk-dominated nova-like CV.  However, as noted by \citet*{Mar94}, there are no strong differences between the optical spectra of dwarf novae and those of quiescent black holes.  In the case of V1, we note the similarity to the broadband spectra of several dynamically-confirmed quiescent BHs \citep{McC03,Fro11,Hyn12}, which could all be fit by blackbodies of temperature 5,000--13,000\,K, with areas much smaller (0.06--1.25\%) than the projected area of the expected truncated quiescent accretion disk (i.e.\ radii of $0.01$--$0.09R_{\sun}$).  This emission was explained as either the impact point of the accretion stream on the disc \citep{McC03}, or as emission from the inner part of the disc, close to the truncation radius \citep{Hyn12}.

The main caveat to this explanation is that the X-ray spectrum of X9 is difficult to reconcile with those of known quiescent black holes, which at luminosities of $10^{33}$\,erg\,s$^{-1}$ are fit by relatively soft power-law spectra with photon indices of $\Gamma\sim2.1$ \citep{Plo13}.  While no evidence for low-energy emission lines close to 0.6\,keV has been observed in any quiescent black hole, the high column densities towards most such systems would have precluded the detection of such a line.  Only three known quiescent systems (XTE J1118+480, A0620$-$00 and MAXI J1659$-$152) have absorption column densities $<5\times10^{21}$\,cm$^{-2}$ \citep{Plo13}, none of which have stringent published constraints on the presence of low-energy emission lines.  The only published quiescent spectrum of sufficiently high quality to investigate the presence of such lines is that of V404 Cyg \citep{Bra07}.  That spectrum was fit with an absorption column of $7$--$9\times10^{21}$\,cm\,$^{-2}$, and clearly excludes a line of similar strength in the range 0.5--0.7\,keV\footnote{The published spectrum shows residuals close to 0.8\,keV, but they are not statistically significant, and a line at this energy has no clear explanation, in contrast to the 0.65\,keV line in X9 that could potentially be explained as O\,{\sc viii}.}.   In summary, the X-ray spectrum, while not fully characteristic of a quiescent black hole, is not better explained by any other known class of source.

\section{The nature of the donor}
\label{sec:nature}

Having identified a black hole as a plausible accretor, we now explore the properties of X9 to help determine the nature of the system.

\subsection{Orbital period}

We reanalysed the optical data of \citet{Edm03b} to check the significance of the marginal 3.5-hour sinusoidal signal and found the false detection probability to be $\gtrsim 10$ per cent, implying a significance of $<2\sigma$.  We also searched for shorter periods than the 1.5--14\,hour range considered by \citet{Edm03b}, finding nothing significant at the $3\sigma$ level.  Thus we have no good photometric constraint on the orbital period.

The measured $V$-magnitude of X9 is in the range 19.8--21.4 \citep{Edm03a,Bec14}, which at 4.57\,kpc implies an absolute magnitude in the range 6.5--8.1.  Should the optical light be attributed to the donor star, this rules out a subgiant or high-mass companion and implies a lower main-sequence donor of spectral class later than mid-K.  Should the optical emission be dominated by the accretion flow (as seems likely given the spectral fitting of \citealt{Kni08}), the donor star would be fainter still.  As determined by \citet{Kni08}, periods shorter than 7\,hr would be required for Roche-lobe overflow of a main-sequence donor star with a sufficiently faint visual magnitude.

At orbital periods longer than a few hours, mass transfer in X-ray binaries is driven by magnetic braking or nuclear evolution of the donor star.  In this regime, there is a well-established positive empirical correlation between orbital period and quiescent X-ray luminosity \citep*{Gar01,McC04,Arm14}, as expected from binary evolution theory in the case where quiescent accretion proceeds via an advection dominated accretion flow \citep{Men99}.  At shorter orbital periods (less than a few hours), angular momentum loss via gravitational wave radiation will take over as the dominant mechanism driving the mass transfer.  This should lead to mass-transfer rates (and hence X-ray luminosities) that increase with decreasing orbital period, implying the existence of a minimum X-ray luminosity at a bifurcation period of $\sim10$\,h \citep{Men99}.  While there is a dearth of known black hole systems at short orbital periods, there is some evidence for a minimum quiescent luminosity \citep{Gal08}, at a level consistent with the best current constraints on the recently-discovered BH candidates in the GC M22 \citep{Str12}.  The recent detection of two black holes with orbital periods $<3$\,h has begun to probe the short-period regime, with MAXI J1659$-$152 \citep{Hom13} and Swift J1357.2$-$0933 \citep{Arm14} both showing higher quiescent luminosities than systems with orbital periods a few times longer (although the uncertain distance prevents a definitive conclusion in the latter case).

At $10^{33}$\,erg\,s$^{-1}$, the X-ray luminosity of X9 is high for a quiescent BH, exceeding that of V404 Cygni (Fig.~\ref{fig:lrlx}), the most luminous known quiescent field BH X-ray binary.  The empirical relationship between orbital period and quiescent X-ray luminosity would then imply either a period that is roughly comparable to the 6.5-day period of V404 Cyg \citep{Cas92}, or else a very short orbital period in an ultracompact system (i.e.\ with a white dwarf or helium star donor).  The former scenario is ruled out by the measured optical emission, although given the scatter on the empirical period-luminosity correlation, we cannot exclude the possibility of a low-mass main-sequence donor with a period $<7$\,h.  Should the system be ultracompact, the absence of He {\sc ii} 1640\,\AA\ line emission in the far-UV spectrum of \citet{Kni08} would suggest a white dwarf rather than a helium star donor.

Although no Galactic ultracompact BH systems have yet been discovered, it is possible to form such systems in a GC environment following an exchange interaction, a long phase of subsequent hardening through stellar encounters, and finally triple-induced mass transfer \citep{Iva10}.  Indeed, the black hole in the extragalactic GC RZ\,2109 in NGC 4472 is believed to be an ultracompact system with a white dwarf donor and an orbital period of $\sim5$\,min \citep{Zep07,Zep08,Gne09} and also shows an excess of soft photons that can be modelled as O\,{\sc viii} emission \citep{Jos15}, so such a scenario would not be wholly unprecedented.  

Given the similarities between the quiescent luminosities of X9 and V404 Cyg, we might expect similar mass transfer rates, i.e.\ $\sim10^{-9}M_{\sun}$\,yr$^{-1}$ \citep{Kin93}.  Such a high mass transfer rate would require the system to be ultracompact, with an orbital period of $\sim20$\,min \citep{vanHaa12}.  While neutron stars in ultracompact systems with such short orbital periods are all persistent X-ray sources \citep{Hei13}, the low X-ray luminosity of X9 implies that it must be a quiescent transient system, which undergoes occasional outbursts due to a disc instability.  Since the critical mass transfer rate for the disc instability model scales with accretor mass as $M_{\rm a}^{0.3}$, then black hole systems will become transient at slightly higher mass transfer rates.  \citet[][see their fig.~8]{vanHaa12} show that the highest mass transfer rate for which a black hole ultracompact system will become transient (a few tens of Myr following the onset of mass transfer) is a few times $10^{-10}M_{\sun}$\,yr$^{-1}$ (assuming a pure He donor), with an orbital period of $\sim25$\,min and a donor of mass $0.02M_{\sun}$.  If the mass transfer rate is to remain approximately similar to that of V404 Cyg, the system should not be significantly older (and with a consequently longer orbital period) than this.

For the disc instability model to operate, the outer disc would need to be below the ionisation temperature of the material, implying that the $\sim0.18R_{\sun}$ optically-thick component fit to the SED by \citet{Kni08} can not extend over the entire disc.  Should the system be ultracompact, then for a 25-minute orbit and a BH mass of $3$--$20M_{\sun}$, the circularization radius should be $1.5$--$3.7\times10^{10}$\,cm \citep[from][fig.~5]{vanHaa12}, so the fitted emission size would be 34--85\% of the circularization radius.  While this might seem uncomfortably large, the physical size is comparable to (albeit slightly larger than) that seen in other quiescent BH X-ray binaries \citep[][as discussed in Section~\ref{sec:bh}]{McC03,Fro11,Hyn12}, and is consistent with typical truncation radii of quiescent discs \citep[a few thousand Schwarzschild radii; see, e.g.][]{Men00}.  While a cooler outer disc might be easier to accommodate in a longer-period, main-sequence system, an ultracompact system remains plausible, particularly given that the non-simultaneity of the data and the time-varying nature of the source permitted only a qualitative fit to the SED.  Quasi-simultaneous, broad-band photometry would be required to properly characterise the size of the emitting region and hence ascertain the viability of this scenario.

Within the disc truncation radius, quiescent black holes host radiatively inefficient accretion flows \citep[e.g.\ advection dominated accretion flows; ADAFs,][]{Nar94}, with a population of hot electrons producing the observed X-ray emission via thermal Comptonisation.  In an ultracompact system, the accreted material would consist of more massive ions than the hydrogen-rich material being accreted by all known quiescent black holes in the field.  The additional gravitational energy liberated could potentially increase the temperature of the particles within the ADAF, thereby affecting the photon index of the power-law and leading to a harder spectrum for X9 in comparison to known quiescent black holes.  However, an in-depth exploration of this issue is beyond the scope of this paper.

\subsection{Constraints on the \ha\ emission}

Since an ultracompact system must have a hydrogen-deficient, degenerate donor star, a test for the presence of \ha\ emission could also help to confirm or refute the ultracompact scenario.  Whereas no Balmer emission would be expected from an ultracompact system, the H$\alpha$ equivalent width (EW) of V404 Cyg was measured by \citet{Cas93} to be $-38.7$\,\AA.  This is roughly as expected from the anticorrelation between X-ray luminosity and EW reported by \citet{Fen09}, for a system of luminosity $10^{33}$\,erg\,s$^{-1}$.

\begin{figure*}
\centerline{\includegraphics[width=\textwidth]{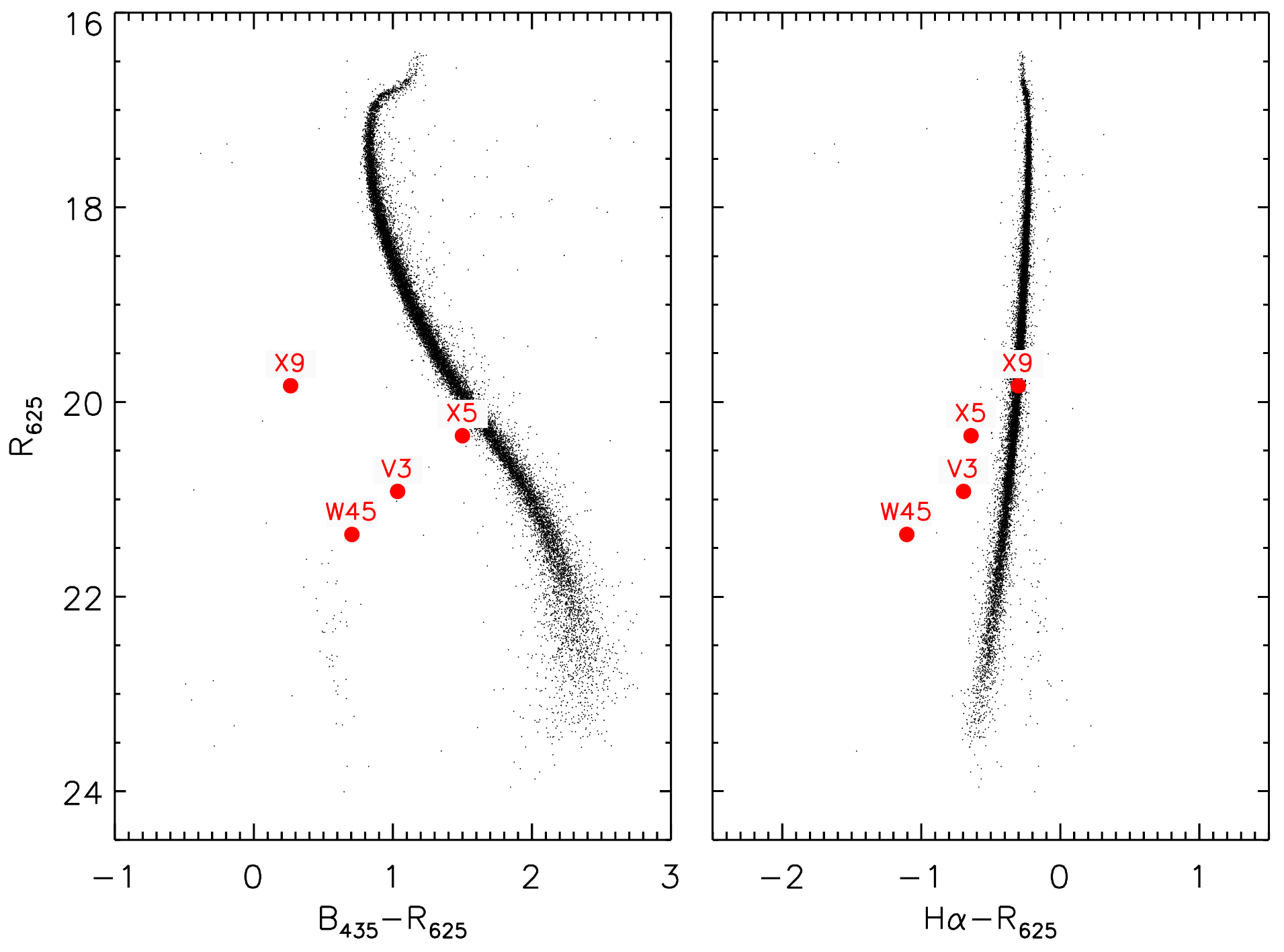}}
\caption{Colour-magnitude diagrams showing the $B_{435}$, $R_{625}$, and H$\alpha$ photometry obtained from the {\em HST} ACS/WFC GO-9281 images of 47 Tuc. The optical counterpart to X9 is shown in red, and the surrounding stars are plotted with black points. For  comparison, we also show (in red) the positions of the optical counterparts to three other {\em Chandra} sources in 47 Tuc: V3 (W27/X10) and W45 (classified as CVs) and X5 (W58; classified as a quiescent NS LMXB). \label{fig:cmd}}
\end{figure*}

\citet{Bec14} used archival {\it HST} data to calculate ($V_{606}-$\ha) and ($V_{606}$-$I_{814}$) colours for X9, which showed an \ha\ excess with an equivalent width of $-125$\,\AA.  However, their $V_{606}$ and $I_{814}$ colours on the one hand, and \ha\ data on the other, were taken in different years, and a comparison of their $V$-band (F606W) data from 2006 March with F555W observations from 1999 July \citep{Edm03a} shows a difference of 1.55 magnitudes, sufficient to render their non-simultaneous ($V_{606}-$\ha) colour meaningless.  

To obtain a better measurement of the H$\alpha$ emission in X9, we analyzed an {\em HST} data set of quasi-simultaneous images of 47\,Tuc in H$\alpha$ and the surrounding continuum. The observations
were taken with the Wide Field Channel (WFC) on the Advanced Camera for Surveys (ACS) under program GO-9281 (PI: J.\,Grindlay). Images were obtained through blue (F435W), red (F625W) and H$\alpha$
(F658N) filters, with total exposure times of 955\,s, 1320\,s, and 7440\,s, respectively.  The observations were split over three visits, on 2002 September 30, October 2, and October 11, with observations being taken in all three filters at each epoch. The data were taken simultaneously with the 2002 {\em Chandra} observations of 47\,Tuc presented in \citet{Hei05}; a full description of the GO-9281 data and results is forthcoming (van den Berg et al.~in preparation).

We retrieved the flat-fielded images, corrected for charge-transfer efficiency losses, from the {\em HST} archive. These images still suffer from significant geometric distortion, which we corrected for
with the DrizzlePac software \citep{Gon12}. The individual images were astrometrically aligned and subsequently combined for each filter using {\em astrodrizzle}, which creates stacked images that are
cleaned of cosmic rays, and that are without geometric distortion. These master images are twice-oversampled to 0\farcs025 per pixel. All images except two 10-s F625W exposures and one 10-s F435W exposure, were included in the stacks.

\begin{figure}
\centerline{\includegraphics[width=\columnwidth]{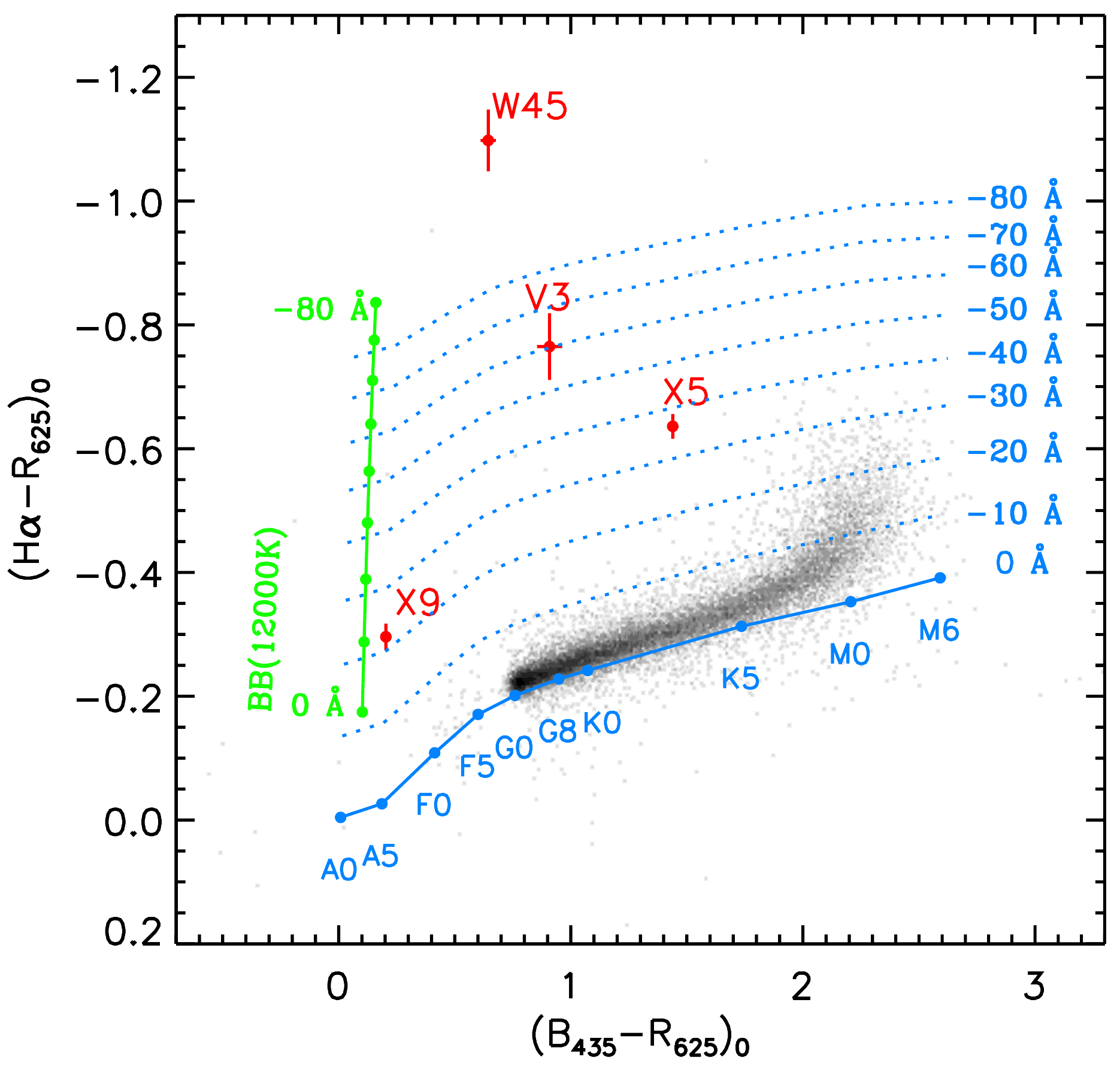}}
\caption{Colour-colour diagram showing in red the optical counterparts to X9 and the three comparison {\em Chandra} sources from Fig.~\ref{fig:cmd}.  Colours of surrounding 47 Tuc stars are shown as black points. The photometry has been dereddened to account for the small extinction towards 47 Tuc ($E(B-V)=0.04$). The expected colours for main-sequence stars of various spectral types are shown in blue, and were computed with the {\tt synphot} package using low-metallicity ($\log Z = -2.5$) Castelli-Kurucz model spectra. The solid blue line is for EW(H$\alpha$)=0\,\AA, and the dotted lines show the effect of adding a gaussian emission line with a FWHM of 25\,\AA~in increments of $\Delta$EW(H$\alpha$)=$-10$\,\AA. In green we show the synthetic colours for a 12,000\,K blackbody model \citep{Kni08} with an added gaussian emission line of different EWs. \label{fig:ccd}}
\end{figure}

Photometry for X9 and the surrounding stars was extracted using the point-spread-function fitting package {\tt DAOPHOT}. The magnitudes were calibrated to the VEGAMAG system using the zeropoints on the
STScI website\footnote{http://www.stsci.edu/hst/acs/analysis/zeropoints}. The position of X9 in the colour-magnitude diagram (Fig.~\ref{fig:cmd}) shows that it is indeed a very blue object, but, in contrast to the findings of \cite{Bec14}, it does not stand out in (H$\alpha-R_{625}$) colour from other stars of similar $R_{625}$ magnitude. Since the average (H$\alpha-R_{625}$) colour of stars varies with ($B_{435}-R_{625}$) colour, we also compared the position of X9 in a colour-colour diagram (Fig.~\ref{fig:ccd}). Unfortunately, X9 is one of the bluest stars in the cluster, so there is a dearth of similarly blue stars.  Instead, an estimate of the H$\alpha$ equivalent width can be made by comparing the unreddened colours of X9 with those expected for the blue component believed to dominate the spectrum of X9 between ultraviolet and infrared wavelengths \citep{Kni08}.

If X9 is an ultracompact binary, a 12,000 K blackbody would be a reasonable approximation of the optical spectrum. The difference between the actual and expected (H$\alpha-R_{625}$) colour then suggests an excess emission in the H$\alpha$ filter that is similar to the effect of adding an H$\alpha$ emission line with an EW of about $-10$\,\AA. In an ultracompact binary, no H$\alpha$ emission is expected. The small inferred excess could point at an incorrect description of X9's continuum emission by the assumed blackbody, or perhaps at the presence of an emission line other than H$\alpha$; for example, the C\,{\sc ii}/O\,{\sc ii} emission line at 6580\,\AA\ in the spectrum of the ultracompact binary 4U 0614+091, which has an EW of $\sim-3$\,\AA\ \citep {Nel04,Bag14}. Since the X-ray luminosity of 4U 0614+091 \citep[$3\times10^{36}$\,erg\,s$^{-1}$;][]{Mig10} is significantly higher than that of X9, the outer disc in X9 is expected to be more optically thin in the continuum.  Since the line emission remains optically thick \citep[e.g.][]{Wil80}, the EW of these C/O lines might be expected to be larger in X9 \citep[as also manifested in the anticorrelation between X-ray luminosity and H$\alpha$ EW;][]{Fen09}.

If X9 is not an ultracompact binary, then the appropriate comparison model should include absorption lines as seen in the spectra of optically thick accretion disks. Comparing the colours of X9 with those for an early A-type star would then suggest an H$\alpha$ EW of about $-20$\,\AA. In this case, part of the excess emission could be explained by a true H$\alpha$ emission line associated with the accretion disk, or part could be emission from the donor star, or perhaps a disc wind. The observed flux at red wavelengths in the SED of X9 could accommodate the signature of a cool ($\sim4000$\,K) secondary \citep{Kni08}, but not that of an A-type main sequence star.  Furthermore, given that X9 is variable and that the data points that make up the SED were not taken simultaneously, such an additional component may not be required.  Regardless, it is clear that there is no strong evidence for an H$\alpha$ excess from X9.

We can also compare our estimate of the H$\alpha$ EW in X9 with that found by \citet{Wan09} for the transitional MSP PSR J1023+0038.  Summing both the red and blue components of the H$\alpha$  line in PSR J1023+0038, \citeauthor{Wan09} found a line flux of  $1.8\times10^{-14}$\,erg\,cm$^{-2}$\,s$^{-1}$.  Combining the $R_{625}$ magnitude of 19.8 with our estimated H$\alpha$ EW for X9 of $-10\,$\,\AA\ and accounting for the difference in source distance \citep[assuming 1.37\,kpc for PSR J1023+0038;][]{Del12}, we find the H$\alpha$ line luminosity of X9 to be a factor of $\sim 7$ times lower than that of J1023+0038, suggesting that X9 has either a smaller disc or less H$\alpha$ emission.  Despite similar absolute magnitudes in the $U$ band \citep{Kni08,Pat14}, PSR J1023+0038 has a redder continuum than X9 (a ($B-R$) colour of 1.0 compared to 0.4 for X9).  A comparison of the optical magnitude of the accreting (disc-dominated) state of PSR J1023+0038 with that in its rotation-powered state suggests that the optical continuum emission is dominated by the disc rather than the donor star in the accretion-powered state \citep{Wan09,Tho05}.  Thus, the redder continuum implies more emission from the outer disc.  Together with the correlation between orbital period and quiescent X-ray luminosity, the anticorrelation between H$\alpha$ EW and quiescent X-ray luminosity \citep{Fen09} implies that shorter-period systems (with a smaller disc) should have a higher H$\alpha$ EW.  Since the larger disc in PSR J1023+0038 has a higher H$\alpha$ EW, we therefore conclude that X9 is likely hydrogen-deficient.

Finally, we note that $\sim70$\% of CVs are found to be \ha\ emitters, with the only magnetic systems found to lack \ha\ emission being polars in the low state \citep{Wit06}.  This provides additional circumstantial evidence against the CV scenario, although we note that the sample of \citeauthor{Wit06} included only a single intermediate polar.

\section{Discussion}
\label{sec:discussion}

X9 is the fourth BH candidate in a Galactic GC discovered via deep radio continuum observations. By design, the clusters selected for our deep radio observations (M22, M62 and 47 Tuc) were all relatively massive \citep[$M>3\times10^5 M_{\sun}$;][]{Pry93,Kim15}. In the absence of significant natal kicks, the large masses imply a significant initial population of BHs, and the long evolutionary timescales make them more likely to retain BHs \citep{Heg14}.  While the absence of any BH candidates in M15 \citep{Str12b} might suggest that other factors are at play, we note that at 10.4\,kpc \citep{Har96}, M15 is over 50\% more distant than the most distant of the other three clusters, M62.  At the distance of M15, the four BH candidates detected to date would have a radio flux density of $\sim8$\,$\mu$Jy\,beam$^{-1}$, which would be $<4\sigma$ in the deepest existing image of the cluster \citep{Str12b}, and hence would remain unremarked.

Given that both recent observational and theoretical work point to a population of BHs in GCs, the absence of any detected X-ray outbursts from such objects in the Galaxy might at first seem surprising. All ten of the bright, transient LMXBs in Galactic GCs \citep{Bah14} have been identified as hosting NSs. This could be taken to suggest that either the ratio of NS to BH LMXBs in GCs is enhanced over that seen in the field, or the properties of BH LMXBs found in GCs differ from those in the field. The likelihood of this can be addressed empirically. Combining the last comprehensive LMXB catalogue \citep{Liu07} with a more up-to-date catalogue of field BH XRBs (B. Tetarenko et al., in prep.), we determined the number of transient systems of different types: 29 LMXBs with dynamically-confirmed BHs or strong BH candidates; 14 LMXBs that are weak BH candidates (and thus could be confused with NS LMXBs); 60 LMXBs with either Type I X-ray bursts or no indication of being a BH (and thus are assumed to be NS LMXBs). Ignoring the harder-to-differentiate weak BH candidates, we estimate the fraction of transient LMXBs in the field that are BHs to be $32.6^{+5.0}_{-4.8}$\%. We can then estimate the probability of observing zero BH LMXBs in GCs given the ten known transient NS LMXBs, assuming the field and GCs share the probability density function of a LMXB containing a BH in the field. We find this null hypothesis probability to be just 2.3\%. While this suggests there may be a difference between the field and GC populations of transient LMXBs, being $<3\sigma$ it cannot be taken as conclusive evidence. We note that this statement remains true even if we classify all the weak BH candidates in the field as BHs.

Furthermore, we note that three of the four cluster BH candidates identified to date could be ultracompact systems (see \citealt{Str12} for M22, and Section~\ref{sec:nature} for 47 Tuc), whereas no such systems are known in the field.  \citet{Kne14} suggest that BHs in short orbital period binary systems could remain undetected due to selection effects.  The correlation between orbital period and peak outburst luminosity \citep{Wu10} implies that the luminosity of short orbital period BHs may never exceed the threshold for radiatively efficient accretion, causing a system to remain in a low-luminosity hard state throughout its outbursts \citep{Mey04}.  The weak irradiation also leads to significantly shorter outbursts, and the duty cycles are similarly reduced to low levels.  The combination of these effects then implies that such outbursts could have been undetectable or overlooked by current or previous All Sky Monitors.  A more sensitive future wide-field monitor such as proposed for LOFT \citep{Mac15} would be needed to detect any such faint outbursts.

Finally, we can make a simple comparison of the relative populations of accreting neutron star and black hole X-ray binaries in the five clusters with deep, published radio imaging \citep[M15, M19, M22, M62, 47 Tuc;][this work]{Str12,Str12b,Cho13}.  At present, we know of three neutron star LMXBs in M15 \citep*{Aur84,Cha86,Hei09}, none in M22 \citep{Web04,Web13}, four in M62 \citep{Poo03,Bah15}, and five in 47 Tuc \citep{Hei03,Hei05b}.  M19 has no existing {\it Chandra} or {\it XMM} data, and the deepest X-ray limit to date is from {\it ROSAT}, with $L_{\rm X}=3\times10^{33}$\,erg\,s$^{-1}$ \citep{Rap94}, which is not sufficiently deep to detect a population of quiescent accreting neutron stars.  Thus we know of 12 accreting neutron stars in our small cluster sample, as compared to four black hole candidates.  Although the sample of clusters is still small, the nature of the black hole candidates is still to be confirmed, and both populations are the subject of poorly-understood selection effects (not least the varying sensitivities achieved in different clusters in the two bands), this implies a ratio in the detectable number of accreting black holes to neutron stars on the order of 1:3.  Given the Galactic globular cluster population of $\sim200$ LMXBs \citep{Hei05b}, this would then imply of order 60 accreting black holes in clusters, which could in principle be detected by combining deep radio and X-ray observations, almost doubling the known population of black hole candidates from the field. 

\section{Conclusions}
\label{sec:conclusions}

We have detected persistent radio emission from the known X-ray source X9 in the globular cluster 47 Tuc, which had previously been identified as an intermediate polar.  The measured radio flux density of $42\pm4$\,$\mu$Jy\,beam$^{-1}$ at 5.5\,GHz implies a radio luminosity of $5.8\times10^{27}$\,erg\,s$^{-1}$, which is significantly brighter than the persistent emission seen from any cataclysmic variable.  The radio emission appears to be roughly constant from 2010 to 2013, as well as within an individual 10-hour observation.  Together with the absence of significant circular polarization, this also allows us to rule out flaring emission from a cataclysmic variable.  Although transitional MSP systems are known to produce flat-spectrum radio emission in their accreting states, they show lower radio luminosities, simpler X-ray spectra, and rather different X-ray variability than does X9.  While we cannot definitively rule out this explanation, we find it unlikely.  Combining our radio detection with archival {\it Chandra} X-ray data, we find that 47 Tuc X9 lies very close to the radio/X-ray correlation for quiescent stellar-mass black holes.  The similarity of its optical-through-ultraviolet spectrum to those of dynamically-confirmed quiescent black holes further supports X9 being a stellar-mass black hole in quiescence, which would make it the fourth black hole candidate detected in a Galactic globular cluster to date.  While the optical magnitude implies that the orbital period should be $<7$\,h (consistent with a lower-main sequence donor star), the high X-ray luminosity of $10^{33}$\,erg\,s$^{-1}$ and the absence of strong H$\alpha$ emission suggest that the system could be ultracompact, with an orbital period of order 25\,minutes.  Differentiating between these two possibilities would require broadband optical spectroscopy.

\section*{Acknowledgments}
JCAMJ thanks Paul Hancock for assistance with the {\sc aegean} fitting package and Joe Callingham for useful discussions regarding CVs.  We thank the anonymous referee for helpful and constructive comments.  The Australia Telescope Compact Array is part of the Australia Telescope National Facility which is funded by the Commonwealth of Australia for operation as a National Facility managed by CSIRO.  This work was supported by the NSF through grant AST-1308124.  JCAMJ is the recipient of an Australian Research Council Future Fellowship (FT140101082), and also acknowledges support from an Australian Research Council Discovery Grant (DP120102393).  COH and GRS acknowledge funding from NSERC Discovery Grants, and COH is also supported by an Ingenuity New Faculty Award and an Alexander von Humboldt Fellowship. This research has made use of NASA's Astrophysics Data System, and also of the Python packages APLpy (an open-source plotting package for Python hosted at {\tt http://aplpy.github.com}) and Astropy \citep[a community-developed core Python package for Astronomy;][]{Ast13}.

\label{lastpage}
\bibliographystyle{mn2e}

\end{document}